\documentstyle[12pt]{article}
\title{ Global Solutions of the Equations of \\ Elastodynamics of
Incompressible Neo-Hookean Materials}

\author{ David G. Ebin
\thanks{Research done at Department of Mathematics, UCLA, Los Angeles, CA,
90024-1555 }}
\date{}
\begin{document}
\maketitle
To appear in Proc. Nat. Acad. Sci., Vol 90, 1993

\section{Introduction}

     Imagine three-space to be filled with an elastic material and let
\mbox{$\eta(t): \bf R^3 \rightarrow \bf R^3$} be the map which takes each
material point from its position at time zero to its position at time $t$.
Then the curve of maps, $t \mapsto \eta(t)$, will completely describe the
motion of the material.

     We shall assume that the material is incompressible, which in
mathematical terms means that $\eta(t)$ must satisfy the equation:
\begin{equation}
J(\eta(t)) \equiv 1 \label{1.1}
\end{equation}
where $J$ means the Jacobian determinant.  We shall further assume that
the material is ``neo-Hookean" or that when deformed by a map $\eta: \bf
R^3 \rightarrow \bf R^3$, it acquires potential energy:
\begin{equation}
V(\eta) = {1\over 2} \int_{\bf R^{3}} (\partial_i \eta^j \partial_i\eta^j
- 3) \ \ \ \ \ {\rm (repeated ~indices ~summed ~from ~1 ~to ~3)} \label{1.2}
\end{equation}

     From (\ref{1.1}) - (\ref{1.2}) one can derive the equations of
motion which $\eta(t)$ must satisfy. (see \cite{E-Sa}).  They are:
\begin{equation}
\ddot{\eta}^i (t)(x) = (\Delta\eta^i (t))(x) +
(\partial_{i} p(t))(\eta(t)(x)) \label{1.3}
\end{equation}
where ``$\,\cdot{}\,$'' means time derivative, ``$\Delta$" means Laplacian,
and $p(t)$
is a function on $\bf R^3$ which is chosen so that (1) will hold, (as
we shall see below).

     It is known \cite{E-Sa} that given initial data $\eta(0)$ and
$\dot{\eta}(0)$, there is a positive $T$ and a unique curve $\eta(t) \ (0
\leq t < T)$ satisfying (1) and (3).\footnote{By ``solution'' we will
always mean smooth or classical solution.  We do not consider weak
solutions in this paper.}  In this announcement, we shall
show that if $\eta(0)$ is sufficiently near the identity map and if
$\dot{\eta}(0)$ is small, then one can take $T = \infty$.  Thus if the
initial deformations are small, the equations have a unique solution for
all time.

     To provide context to this theorem we now discuss some related results,
each of whose proofs contains ideas that we will use for our proof below.
First we consider the case of compressible elastic media.
If the potential energy is as in
(\ref{1.2}), the resulting equation is the (linear) wave equation, so of course
solutions persist for all time.  However if $V(\eta)$ is not quadratic, the
resulting equation is not linear, so it is possible that solutions will
develop singularities.  F. John \cite{J1} \cite{J2} analyzes this
situation in considerable detail for homogeneous isotropic materials.  He
finds that for this case the situation is essentially the same as for the
simpler case of a single
quasi-linear wave equation, which has been studied over many years by
Klainerman, John, et al. (see \cite{J3} for a survey).
Thus we proceed with a brief discussion of quasi-linear wave equations
(some of which we shall need for the proof of our theorem).

     Let $f = f(x^0,x^1, x^2, x^3)$ (identifying $t$ with $x^0$) and let
$f'$ and $f''$ denote
respectively its first and second order partial derivatives.  Let $f$
satisfy:
\begin{equation}
\Box f = F(f', f'') \label{1.4}
\end{equation}
where $\Box = \partial_0^2 - \partial_{1}^2 - \partial_{2}^2 -
\partial_{3}^2, \ F$ is linear in $f''$, and: $$F(f', f'') =
O(|f'|(|f'| + |f''|))$$
near zero.  Then John and Klainerman prove the following:
\cite{J-K}

{\bf Theorem 1}.  {\it Let $f^0(x)$ and $f^1(x)$ be smooth functions on
$\bf R^3$ with compact support.  Then there exist positive constants $A,
B$, and $\varepsilon_0$ such that if $\varepsilon < \varepsilon_0$,
then equation {\rm (\ref{1.4})} with initial data
$f(0,x) = \varepsilon f^0(x)$ and $\partial_t f(0,x)
= \varepsilon f'(x)$ has a unique solution for $0 \leq t < T$ with $T \geq
Ae^{B/\varepsilon}$.}

     Thus a quasi-linear wave equation has solutions for a very long time
if the data are small. Moreover as John points out, (\cite{J3} p. 27) this
result is sharp.  That is letting: $$F(f', f'') =
\left({-2\partial_{t}f\over 1-2\partial_{t}f}\right)\Delta f$$
and choosing $f^1$ with support in the unit ball,
we find that (\ref{1.4}) does not have a smooth solution on $[0,T)$ for
$T>e^{1/L}$ where:
\begin{eqnarray}
%T &>& 2 e^{1/L} \ \ {\rm where} \nonumber \\
L &=& {3\over 64\pi} \int_{\bf R^{3}} f^1(x) - (f^1(x))^2 dx
\nonumber
\end{eqnarray}

     Theorem 1 is proven by looking at various weighted spatial $L^2$
norms of $f$ and its derivatives and showing that they obey certain
differential inequalities.  In particular one shows that if $N(t)^2$ is a
certain sum of terms of type:
\begin{equation}
\int_{\bf R^{3}}|(x^\alpha)^l (\partial_\beta)^kf(t,x)|^2 dx \ \ \alpha,
\beta = 0,1,2,3, \ \ \ell \leq k \nonumber
\end{equation}
with $t$ identified with $x^0$, and $x=(x^1,x^2,x^3)$, then one gets an
inequality of type:
\begin{equation}
{d\over dt} N(t) \leq {\rm const.} (1+t)^{-1} N(t)^2 \label{1.6}
\end{equation}
Using this and the fact that $\int_0^T(1+t)^{-1}dt = \log(1+T)$ one gets
the necessary estimates to prove Theorem 1.

     On the other hand, the sharpness of Theorem 1, or the necessary
development of singularities is shown by looking at spherical means of
functions depending on $f$ and its derivatives.
Using these one constructs a function $W(t)$ which obeys:
\begin{equation}
{d\over dt} W(t) \geq {\rm const.} (1+t)^{-1} W(t)^2 \ \ \ \ \ \
{\rm (the ~reverse ~of ~(\ref{1.6}))}
\label{1.7}
\end{equation}
and from this gets an upper bound on the time of existence of a smooth
solution.

     John shows that for compressible materials, the situation is the same
as for solutions of (\ref{1.4}).
He uses the displacement $u(t,x) = \eta(t)(x) - x$ and notes that it satisfies
an equation:
\begin{equation}
\partial_t^2 u = c_1 \Delta u + c_2 \nabla {\rm div}(u) + F(u', u'')
\label{1.8}
\end{equation}
where ``$\nabla$'' means gradient, ``div'' means divergence, $F$
is like the corresponding function in (\ref{1.4}) and $c_1$ and $c_2$
are positive constants.  For (\ref{1.8}) one can get estimates like those of
(\ref{1.4}) and thus find that given initial data of compact support,
solutions exist as
in Theorem 1. (see \cite{J1})  Here too John shows that the result is
sharp. (see \cite{J2})  Taking spherically symmetric data one gets a solution
of
the form:
\begin{equation}
u(t,x) = \varphi(t,|x|)x \nonumber
\end{equation}
where $\varphi(t,|x|)$ satisfies a quasi-linear wave equation.
Certain combinations of second derivatives of $\varphi$ obey an inequality
like (\ref{1.7}) and thus $\varphi(t,|x|)$ remains smooth for only a finite
time.

     For incompressible elastic materials all results that we know of are
for a finite time which depends on the data.  \cite{E-Sa} proves such a
result for materials filling $\bf R^n$, and \cite{E-Si} and \cite{H-R}
show a similar result for initial-boundary value problems.  Furthermore, we
do have examples where singularities develop. (\cite{E-Si}, section 7)

     Another relevant case is that of
incompressible fluids (the case $V(\eta) \equiv 0$) for which there is a
much larger
literature.  First there is the old result of Wolibner \cite{W}.  It uses
estimates based on conservation of vorticity to get existence of solutions
for all time for fluids in $\bf R^2$.  The same principle is used by Kato
\cite{K} and Yudovi\v{c} \cite{Y} who prove existence for all time for
fluids in two-dimensional domains with boundary. This analysis is further
extended by
Beale, Kato and Majda \cite{B-K-M} who show that existence of solutions for
all time is
guaranteed in $\bf R^3$ provided the vorticity of the flow obeys a
certain bound.

     The proof of our result uses a combination of all of the above techniques.
We first consider the equation in terms of the displacement.  Applying the
curl (or exterior derivative) to that, we get a quantity somewhat
analogous to the vorticity of a fluid, which satisfies an equation similar
to (\ref{1.4}).  Using norms like $N(t)$ above we get an inequality stronger
than (\ref{1.6}), and from that we get existence of solutions for all time.

     We would like to generalize our proof so that it covers incompressible
materials which are not neo-Hookean, but have not been able to do so as yet.

\section{Definitions and Statement of the \mbox{Theorem}}

     We shall be concerned with functions and vector fields on $\bf R^3$,
and we shall use the exterior derivative, ``$d$", on either of these.
Thus for a function $p, \ dp$ is the same as $\nabla p$ and for a vector
field $u, \ du$ is the same as the curl of $u$.  Also we will denote by
$\delta$ the formal adjoint of $d$ so that $\delta u$ is the negative of
the divergence of $u$.

     Furthermore if $L$ is any differential operator and $\eta: \bf R^3
\rightarrow \bf R^3$ is a diffeomorphism, we shall define $L_\eta$ by:
\begin{equation}
L_\eta f = (L(f \circ \eta^{-1})) \circ \eta. \label{2.1}
\end{equation}
$If \ \eta(t)$ is a curve of diffeomorphisms, a direct calculation yields:
\begin{equation}
{d\over dt} (L_{\eta(t)} f) = [v \cdot \nabla, L]_{\eta(t)} f \label{2.2}
\end{equation}
where [~,~] denotes the commutator and $v$ is the vector-field defined by
\begin{equation}
\dot\eta(t) (x) = v(\eta(t)(x)). \label{2.3}
\end{equation}
Finally for functions on space-time we define the operators:
\begin{eqnarray}
\Gamma_{i} &=& \partial_{i} \quad i = 0 \cdots 3 \nonumber \\
\Gamma_{3+j} &=& x^k \partial_{i} - x^i \partial_{k} \ {\rm where}~ \{ijk\} \
{}~{\rm ~is ~an ~even ~permutation ~of}~ \{123\} \nonumber \\
\Gamma_{6+k} &=& t\partial_{k} + x^k\partial_0 \quad k = 1\cdots 3
\nonumber \\
\Gamma_{10} &=& t\partial_0 + \sum_{i=1}^3 x^i \partial_{i} \nonumber
\end{eqnarray}

Also for any multi-index $\alpha = (\alpha_0 \cdots \alpha_{10})$ we define
the higher order operator:
\begin{equation}
\Gamma^\alpha = \Gamma_0^{\alpha_0} \Gamma_1^{\alpha_{1}} \cdots
\Gamma_{10}^{\alpha_{10}}. \nonumber
\end{equation}

$\Gamma_0$ to $\Gamma_3$ of course commute with the wave operator $\Box$,
and since $\Gamma_4$ to $\Gamma_9$ are infinitesimal generators of the
Lorentz group, they commute with $\Box$ also.  However $\Gamma_{10}$
generates similarity transformations and for it we have the identity:
\begin{equation}
[\Gamma_0, \Box] = -2\Box. \label{2.4}
\end{equation}
Using the $\Gamma_p$ we define certain $L^2$ norms on functions of space
time which depend on time and are stronger than the Sobolev norms.  We
write:
\begin{equation}
\|f\|_{LS,s}^2 = \sum_{|\alpha|\leq s} \int_{\bf R^{3}} |\Gamma^\alpha
f(t,x)|^2 dx \nonumber
\end{equation}
($LS$ stands for Lorentz-similarity).  We shall also need supremum norms
for functions and their derivatives and for this we write:
\begin{eqnarray}
\|f\|_{C^{k}} &=& \sum_{|\alpha|\leq k} \  \sup_{x\in \bf R^3} |D^\alpha f(x)|
\nonumber \\
{\rm where} \qquad D^\alpha &=& \partial_{1}^{\alpha_{1}}
\partial_{2}^{\alpha_{2}} \partial_{3}^{\alpha_{3}}. \nonumber
\end{eqnarray}

      Having introduced the necessary notation, we can now discuss our
basic equation (\ref{1.3}).  We introduce $q(t)(x) = p(t) (\eta(t)(x))$
and then write the equation as:
\begin{equation}
\ddot\eta = \Delta\eta + d_\eta q. \label{2.5}
\end{equation}
We would like to show that $q$ is determined by $\eta$ and its
derivatives, and to do so we first take a time derivative of equation (1).
This gives:
\begin{equation}
\delta_\eta (\dot\eta) = 0. \label{2.6}
\end{equation}
Then applying a second time derivative and using (\ref{2.2}) gives:
\begin{equation}
\delta_\eta(\ddot\eta) - [v \cdot \nabla, \delta]_\eta \dot\eta = 0.
\label{2.7}
\end{equation}
where $v$ is defined as in (\ref{2.3}).
Also a direct calculation yields:
\begin{equation}
[v\cdot \nabla, \delta]_\eta \dot\eta = {\rm tr}((D_\eta\dot\eta)^2) \
{\rm where ~``tr" ~means ~trace ~of ~a ~matrix}. \label{2.8}
\end{equation}

     Now applying $\delta_\eta$ to (\ref{2.5}) and using (\ref{2.7}) and
(\ref{2.8}) we find:
\begin{equation}
{\rm tr}((D_\eta\dot\eta)^2) = \delta_\eta \Delta\eta - \Delta_\eta q.
\label{2.9}
\end{equation}
Thus given $\eta(t)$ and $\dot\eta(t)$ we get $q(t)$ as the solution of
(\ref{2.9}) and it follows that the right side of (\ref{2.5}) is a function
of $\eta$ and $\dot\eta$.  Our main theorem is:

{\bf Theorem 2}:  {\it Given $\eta_0$ and $\eta_1$ smooth vector fields
on $\bf R^3$ such that $\eta_0 - Id$ and $\dot\eta_1$ are rapidly
decreasing; there exists a maximal $T > 0$ and a unique curve $\eta(t) \
(0 \leq t < T)$ which satisfies \rm{(\ref{1.1})} and \rm{(\ref{2.5})}
 ($q$ defined as in \rm{(\ref{2.9})}) such
that $\eta(0) = \eta_0$ and $\dot\eta(0) = \eta_1$.  Furthermore there
exists $\varepsilon > 0$ such that if $\|\eta_0 - Id\|_{LS,5}$ and
$\|\eta_1\|_{LS,4}$ are less than $\varepsilon$, then $T = \infty$.
}

\section{Outline of Proof}

     We let $u(t,x) = \eta(t)(x) - x$ be the displacement as defined in
\S1.  Then (\ref{2.5}) is equivalent to:
\begin{equation}
\Box u = d_{\eta} q. \label{3.1}
\end{equation}
Also since $\eta(t)(x) = x + u(t,x)$, we find that the equation (1) is
equivalent to:
\begin{equation}
\delta(u) = Q(Du) + \det (Du) \label{3.2}
\end{equation}
where
\begin{equation}
Q(Du) = \sum_{i=1}^3 \det \left(\begin{array}{ll} \partial_i u^i
&\partial_i u^{i+1} \\ \partial_{i+1} u^i &\partial_{i+1} u^{i+1}
\end{array}\right) {\rm ~(indices ~taken~ mod ~3)} \nonumber
\end{equation}
and ``$\det$'' means determinant.  Thus (\ref{3.1}),(\ref{3.2}) is equivalent
to (\ref{1.1}), (\ref{1.3}) and
from \cite{E-Sa} we have the following short time existence theorem.

     {\bf Theorem 3}:  {\it Given $u^0 \in H^4(\bf R^3,\bf R^3)$ and
$u^1 \in H^3(\bf R^3, \bf R^3)$ such that $\delta(u^0) = Q(Du^0) +
J(u^0)$ and $\delta(u^1) = 0$, there exists $T > 0$ and a unique solution
of $u(t,x)$ (\ref{3.1})-(\ref{3.2}) for $0 \leq t < T$ such that
$u(0,x) = u^0(x) \ \
\partial_t u(0,x) = u^1 (x)$.  Furthermore if $u^0 \in H^{s+1}(\bf R^3,
\bf R^3)$ and $u^1 \in H^s(\bf R^3, \bf R^3)$ then $u \in \cap_{k=0}^s
C^k([0,T), \ H^{s+1-k}(\bf R^3, \bf R^3))$.}

     For our estimates we will need the $\|~\|_{LS}$ norms, which are
examples of the norms $N(t)$ of \S1.  Hence we need to know that $u(t,x)$
goes to zero rapidly as $|x| \rightarrow \infty$.  Thus we introduce
$H^{s,\ell}$ norms:
\begin{equation}
\|f\|_{s,\ell}^2 = \int_{\bf R^{3}} |D^sf(x)|^2(1+|x|^2)^{\ell} dx.
\nonumber
\end{equation}
A slight modification of the proof of \cite{E-Sa} shows that in Theorem
3, $H^s$ can be replaced by $H^{s,\ell}$ for any positive integer
$\ell$.  Thus norms of the form $N(t)$ will be finite for solutions $u(t,
x)$.

     To prove the last statement of Theorem 2 it suffices to show that $u$
remains bounded as $t$ gets large.  To do this we will derive an equivalent
equation for $du$ and show that solutions of that equation must remain
bounded.

     To derive an equation for $du$ we first apply $d_\eta$ to (\ref{3.1})
getting:
\begin{equation}
d_\eta \Box u = 0. \label{3.4}
\end{equation}

     However, by direct calculation we find that for any $v: \bf R^3
\rightarrow \bf R^3$,
\begin{equation}
d_\eta v = A((D\eta)^{-1}Dv) \label{3.5}
\end{equation}
where $(D\eta)^{-1}$ denotes the matrix inverse of $D\eta$ and $A$
applied to a matrix means the matrix minus its transpose.  Thus
(\ref{3.4}) can be written:
\begin{equation}
A((D\eta)^{-1} \Box Du) = 0. \label{3.6}
\end{equation}

     Since $D\eta = Id + Du$ (and therefore $(D\eta)^{-1} - Id =
-Du(D\eta)^{-1})$ we can rewrite (\ref{3.6}) as:
\begin{eqnarray}
A(\Box Du) &=& A(Du(D\eta)^{-1} \Box Du) \nonumber \\
\noalign{\hbox{or}}
\Box du &=& A(Du(D\eta)^{-1} \Box Du) \label{3.7}
\end{eqnarray}

     Using LS norms we will derive estimates for (\ref{3.7}) which
show that it has solutions for all time.

   To analyze (\ref{3.7}) we would like to eliminate the highest order terms on
the right side -- namely $\Box Du$, and to do this must first express $Du$
as a function of $du$.  Letting $\omega = du$ and using (\ref{3.2}),
we find that $u$ satisfies the elliptic system:
\begin{equation}
du = \omega, \ \ \delta u = Q(Du) + \det (Du). \label{3.8}
\end{equation}

     {\bf Proposition 4}.  {\it There exists a ball $B$ about zero in
$H^3(\bf R^3, \bf R^3)$ such that given $\omega \in H^2(\bf R^3, \bf
R^3)$ there is at most one $u$ in $B$ which satisfies (\ref{3.8}).}

     {\it Proof}.  Use the fact that $u$ is determined by $du$ and
$\delta u$ plus
the fact that $Q(Du)$ and $\det (Du)$ are $O(|Du|^2)$.

     Proposition 4 defines a map $G(\omega) = Du$ where $u$ is the
solution of (\ref{3.8}).  Thus (\ref{3.7}) can be written:
\begin{equation}
\Box du = A(Du(D\eta)^{-1} \Box G(du)). \label{3.10}
\end{equation}

     We proceed to analyze $\Box G(du)$, our first step being to compute:
$$\partial_\alpha G(du), \ \ \alpha = 0,\ldots,3$$

     Applying $\partial_\alpha$ to (\ref{3.8}) we get:
\begin{equation}
d \partial_\alpha u = \partial_\alpha \omega \qquad \delta \partial_\alpha
u = Q'(Du) D\partial_\alpha u + {\rm det} '(Du) D\partial_\alpha u \label{3.11}
\end{equation}
where $Q'$ and $\det'$ are the derivatives of $Q$ and $\det$.  Thus
$\partial_{\alpha} G(\omega) = D\partial_\alpha u$ where $\partial_{\alpha} u$
is the solution of (\ref{3.11}).  We write
$D\partial_\alpha u = G'(\omega)\partial_\alpha \omega$.

     Applying $\partial_{\alpha}$ again we find that $\partial_{\alpha}^2 u$
satisfies:
\begin{eqnarray}
d \partial_{\alpha}^2 u &=& \partial_{\alpha}^2 \omega  \nonumber \\
\delta \partial_{\alpha}^2 u &=& Q' (Du) D\partial_{\alpha}^2 u +
{\rm det}'(Du) D \partial_{\alpha}^2 u \label{3.12a} \\
& & + Q''(Du) (D\partial_{\alpha} u, D\partial_{\alpha} u) + {\rm det}'' (Du)
(D\partial_{\alpha} u, D\partial_{\alpha} u) \nonumber
\end{eqnarray}
where $Q''$ and ${\rm det}''$ are second derivatives of $Q$ and $\det$.
Thus we can write $\partial_\alpha^2 u = c_\alpha + b_\alpha$ where $c_\alpha$
solves:
\begin{equation}
d c_{\alpha} = \partial_\alpha^2 \omega \ \ \delta c_\alpha = Q'(Du) D
c_\alpha + {\rm det}'(Du) Dc_\alpha \label{3.12b}
\end{equation}
and $b_\alpha$ solves:
\begin{eqnarray}
d b_\alpha & = &0 \nonumber \\
\delta b_\alpha & = & Q'(Du) Db_\alpha + {\rm det}'(Du) Db_\alpha
\label{3.13}\\  & & + Q''(Du) (D\partial_\alpha u, D\partial_\alpha u) +
{\rm det}''(Du)(D\partial_\alpha u, D\partial_\alpha u). \nonumber
\end{eqnarray}
In particular $D c_\alpha = G'(\omega) \partial_\alpha^2 \omega$.  Then summing
over $\alpha$ we get:
\begin{eqnarray}
\Box G(\omega) &=& D(c_0 - \sum_{i=1}^3 c_i) + D(b_0 - \sum_{i=1}^3 b_i)
\label{3.14} \\
&\stackrel{def}{=}& DC + DB \nonumber
\end{eqnarray}
Here
\begin{equation}
DC = G'(\omega)( \Box du) \label{3.15}
\end{equation}
Also combining (\ref{3.14}) with (\ref{3.10}) we get:
\begin{equation}
\Box du - A(Du(D\eta)^{-1}DC) = A(Du(D\eta)^{-1}DB) \label{3.16}
\end{equation}
so using (\ref{3.15}) we get:
\begin{equation}
\Box du - A(Du(D\eta)^{-1}G'(w)\Box du) = A(Du(D\eta)^{-1}DB). \label{3.17}
\end{equation}

     The second term on the left side of (\ref{3.17}) can be thought of as
a linear operator
applied to $\Box du$ and if $Du$ is small it will have small norm.  Hence
the map $\Phi(\Box du) = \Box du - A(Du(D\eta)^{-1} G'(du) \Box du)$
will be a linear map which is near the identity.
Thus $\Phi$ will be invertible
and we can write:
\begin{equation}
\Box du = \Phi^{-1}(A(Du(D\eta)^{-1}DB). \label{3.18}
\end{equation}

     Our theorem can now be proven by getting estimates for the solution
of (\ref{3.18}).

     We note that $DB$, which is a sum of $Db_\alpha$, depends on
$D\partial_\alpha u$ in a bi-linear fashion.  Therefore, letting
$\partial Du$ denote the four partials of $Du$, we get:
\begin{equation}
Du(D\eta)^{-1} DB = O(|Du||\partial Du|^2) \nonumber
\end{equation}
and since $\Phi$ goes to the identity as $Du \rightarrow 0$, we find:
\begin{equation}
\Phi^{-1} (A(Du(D\eta)^{-1}DB) = O(|Du||\partial Du|^2)
\label{3.19}
\end{equation}
also.

     Let $E$ denote the energy associated with the wave equation, i.e.
\begin{equation}
E(g) = {1\over 2} \int_{\bf R^{3}} (|Dg|^2 + |\partial_tg|^2).
\label{3.20}
\end{equation}
Then ${d\over dt} E(g) = \int_{\bf R^{3}} (\Box g)(\partial_t g)$.
Therefore:
\begin{equation}
{d\over dt} E(du) = \int \langle \Phi^{-1} (A(Du(D\eta)^{-1} DB)),
\partial_t du\rangle.
\end{equation}
Then applying $\Gamma^\alpha$ to (\ref{3.18}) (with $|\alpha| \leq 4$) and
using
(\ref{3.20}) and the commutation properties of $\Box$ and $\Gamma^\alpha$,
one gets:
\begin{eqnarray}
{d\over dt} \:  E(\Gamma^\alpha du)& = &  \int \langle \Gamma^\alpha
\Phi(A(Du(D\eta)^{-1}DB)), \Gamma^\alpha \partial_t du\rangle \label{3.21}
\\
& &\leq K \|\partial_t du\|_{LS,4}  \, \|\partial du\|_{LS,4} \,
\|Du\|_{C^{3}}^2 \nonumber
\end{eqnarray}
for some constant $K$.

     But Klainerman's inequality (see \cite{J3}, appendix 2) combined with
the estimates for $Du = G(du)$ tells us that:
\begin{equation}
\|Du\|_{C^{3}}^2 \leq {K\over (1+t)^2} \sum_{|\alpha|\leq 4}
E(\Gamma^\alpha du). \label{3.22}
\end{equation}
Let $E_4 (du) = \sum_{|\alpha|\leq 4} E(\Gamma^\alpha du)$.  Then
$E_4(du)$ bounds $\|\partial du\|_{LS,4}^2$, so from (\ref{3.21})
and (\ref{3.22}) we find that there is a constant $K$ such that:
\begin{equation}
{d\over dt} E_4(du) \leq K (1+t)^{-2} E_4 (du)^2. \label{3.23a}
\end{equation}
But integrating this inequality we find that if $E_4 (du) (0) < {1\over
K}$, then:
\begin{equation}
E_4(du)(t) \leq {E_4(du)(0)\over 1-KE_4(du)(0)(1-{1\over 1+t})}
\label{3.23b}
\end{equation}
for all positive $t$.

     From this it follows that if $du$ (and its derivatives) is small at
time zero, then it must remain bounded for all time.  Thus the solution of
(\ref{3.18}) remains bounded, and hence the solution of (\ref{3.1}) must
remain bounded as well.

{\sc SUNY Stony Brook, NY 11794-3651} \\
e-mail: ebin@math.sunysb.edu
\end{document}